\documentclass[reprint,amsmath,amssymb,aps,superscriptaddress,prc,floatfix]{revtex4-1}
\usepackage{amsmath,amsfonts}
\usepackage{graphicx}
\usepackage{longtable}
\usepackage{bm}
\usepackage{hyperref}
\hypersetup{colorlinks=true,breaklinks,linkcolor=red,citecolor=blue}

\begin{document}
\title{Comparison of nuclear hamiltonians using spectral function sum rules}

\author{A. Rios}
\affiliation{Department of Physics, Faculty of Engineering and Physical Sciences, University of Surrey, Guildford, Surrey GU2 7XH, United Kingdom}

\author{A. Carbone }
\affiliation{Institut f\"ur Kernphysik, Technische Universit\"at Darmstadt, 64289 Darmstadt, Germany}
\affiliation{ExtreMe Matter Institute EMMI, GSI Helmholtzzentrum f\"ur Schwerionenforschung GmbH, 64291 Darmstadt, Germany}

\author{A. Polls}
\affiliation{Departament d'Estructura i Constituents de la Mat{\`e}ria and Institut de Ci{\`e}nces del Cosmos, Universitat de Barcelona, Avinguda Diagonal 647, E-8028 Barcelona, Spain}

\date{\today}

\begin{abstract}
\begin{description}
\item[Background] 
The energy weighted sum rules of the single-particle spectral functions provide a quantitative understanding of the fragmentation of nuclear states due to short-range and tensor correlations. 
\item[Purpose] 
The aim of this paper is to compare on a quantitative basis the single-particle spectral function generated by different nuclear hamiltonians in symmetric nuclear matter using the first three energy-weighted moments.
\item[Method] 
The spectral functions are calculated in the framework of the self-consistent Green's function approach at finite temperature within a ladder resummation scheme. We analyze the first three moments of the spectral function and connect these to the correlations induced by the interactions between the nucleons in symmetric nuclear matter. In particular, the variance of the spectral function is directly linked to the dispersive contribution of the self-energy. The discussion is centered around two- and three-body chiral nuclear interactions, with and without renormalization, but we also provide results obtained with the traditional phase-shift-equivalent CD-Bonn and Av18 potentials. 
\item[Results] 
The variance of the spectral function is particularly sensitive to the short-range structure of the force, with hard-core interactions providing large variances. Chiral forces yield variances which are an order of magnitude smaller and, when tamed using the similarity renormalization group, the variance reduces significantly and in proportion to the renormalization scale. The presence of  three-body forces does not substantially affect the results.
\item[Conclusions] 
The first three moments of the spectral function are useful tools in analysing the importance of correlations in nuclear ground states. In particular, the second-order moment provides a direct insight into dispersive contributions to the self-energy and its value is indicative of the fragmentation of single-particle states.
\end{description}
\end{abstract}

\maketitle

\section{Introduction}

The microscopic understanding of infinite nuclear matter has always been a central issue in theoretical nuclear physics \cite{baldo1999}. In an \emph{ab initio} philosophy, the starting point is a nuclear hamiltonian with two-body nuclear forces (2NFs) that accurately describe the two-nucleon system, including scattering phase shifts and deuteron properties. 
After the interaction is chosen, the quantum many-body problem is solved within a given theoretical approach to produce plausible conclusions about nuclear matter properties \cite{Polls2000,Dickhoff2004a}. The comparison with known properties of nuclear systems around saturation can be used to draw conclusions on the interaction itself \cite{Machleidt1989}. General statements might require a thorough comparison between different many-body methods \cite{Baldo2012}, but insight can already be gained within a single approach. For instance, independently of the many-body formalism, it is well known that the majority of realistic 2NFs fail to reproduce the saturation properties of nuclear matter \cite{Polls2000,Dewulf2003,Li2006}. It is generally accepted that this limitation is associated to the lack of three-nucleon forces (3NFs)  and their important role in dense nuclear systems \cite{Hammer2013}. The bulk saturation properties therefore provide a relevant benchmark for nuclear interactions that are used in \emph{ab initio} finite nucleus calculations \cite{Ekstroem2015}. In this contribution, we want to go beyond bulk properties and explore whether single-particle information, and particularly the fragmentation of single-particle strength, can be linked back to nuclear interactions. 

When 3NFs are included, first principle calculations should be based on a Hamiltonian where both two- and three-body forces are  defined within the same model space. In that sense, chiral effective field theory ($\chi$EFT) attempts to provide a consistent picture of both 2NFs and 3NFs, see Refs. \cite{Epelbaum2009,Machleidt2011} for recent reviews. Within this approach, nuclear interactions are built in terms of nuclear and pionic degrees of freedom. 2NFs and 3NFs are constructed according to a consistent power counting, which can be systematically improved by including higher orders in an expansion of low momenta over a large chiral-breaking scale. 3NFs, for instance, appear at next-to-next-to-leading order (N2LO) in the chiral expansion. In the recent past, 3NFs have been consistently included in several many-body schemes, both in finite and infinite nuclear systems \cite{Hammer2013,Hebeler2015} 

Chiral forces are naturally cut-off at high momenta by regulators, as expected from an effective field theory perspective. As a consequence, one expects them to be ``softer" than traditional phenomenological phase-shift equivalent forces like 
CD-Bonn \cite{Machleidt1989,Machleidt2001} and Argonne v18 (Av18) \cite{Wiringa1995}. The meaning of ``softer" here is not particularly well defined. Naively, one expects that soft forces generate smaller high-momentum components, for instance. Further, there are also arguments that connect softness to the size and extent of off-diagonal matrix elements in relative momentum \cite{Bogner2010}. In any case, both soft chiral forces and hard phenomenological interactions are able to reproduce  to a large extent the Nijmegen phase-shift database \cite{stoks1993}. The application of renormalization techniques to eliminate higher and off-diagonal momenta  by means of the similarity renormalization group (SRG) \cite{Bogner2010} has provided even softer effective forces, which have been also  extensively used  in infinite \cite{Hebeler2010a,Hebeler2011,Carbone2013b,Drischler2016}  and finite nuclear systems\cite{Hagen2010,Roth2012,Barbieri2013,Hergert2013,Hagen2015,Simonis2016}. 

Chiral nuclear forces are not only physically appealing but also do a good job in 
evaluating the   binding energies of  nuclear systems \cite{Hammer2013}. However, when calculating 
certain single particle properties, they are naturally cut-off by the use of regulators - particularly if these are implemented in momentum space. 
One expects these soft interactions to be very useful to calculate low-energy nuclear structure observables, but they can not provide information on the high-momentum contents of the nuclear wave function \cite{Rohe2004,Arrington2011}. 

In the past, several quantitative many-body tools have been developed  to directly address nucleon-nucleon correlations, i.e. mainly the short-range repulsion,
in the nuclear wave function. The Brueckner hole-line expansion includes a 
non-perturbative resummation of particle-particle ladder diagrams which constitutes  the basis of  the traditional Brueckner--Hartree--Fock approach \cite{baldo1999,Polls2000}.
Within the variational approach, correlations induced by the interactions can be directly incorporated into the nuclear wave function. The calculation of the expectation value of the Hamiltonian on these correlated basis functions (CBF) requires unique procedures based on the so-called Fermi-hypernetted chain approach \cite{Lovato2012}.  Recently, quantum Monte Carlo techniques with  local and non-local versions of the chiral nuclear interactions have also tackled short-range correlations directly \cite{Gezerlis2013,Roggero2014,Wlazlowski2014,Lynn2016}. 

Here, we focus on a different approach which is directly linked to the fragmentation of single-particle states by means of the spectral function. 
The implementation of the self-consistent Green's function (SCGF) approach to the  solution of the nuclear many-body problem within a ladder approximation, i.e. particle-particle and hole-hole
intermediate diagrams,
provides direct access to the microscopic properties related to the single-particle propagator \cite{Dickhoff2004a,Rios2009a,Rios2009b,Soma2008,Rios2014,Carbone2013a,Carbone2014}.
These include self-energies, spectral functions, and momentum distributions, from which one can also calculate bulk properties. These single-particle properties, are consistently determined together with the effective nucleon-nucleon interaction in the medium. The method therefore establishes a very close connection between single particle properties and the binding energy of the system. 

Energy-weighted sum rules provide a useful tool to quantitatively characterise the single-particle spectral function. These sum rules are well established in the literature and have been previously used in infinite matter \cite{Polls1994,Frick2004,Rios2006}, finite nuclei \cite{Duguet2012,Duguet2015} and the homogeneous electron gas \cite{Vogt2004}.  The analysis of the sum rules provides useful insights into the numerical accuracy of the many-body approach. Sum rules are also helpful in quantifying the importance of the high momentum components in the nuclear wave function. 
The aim of this paper is to analyze the fragmentation of single-particle strength in symmetric nuclear matter, using the information contained in the energy-weighted sum rules of the spectral function. 

This paper is organized as follows. Section \ref{sec:sumrules} provides a short review of the energy weighted sum rules for the single-particle spectral functions. Numerical results for symmetric nuclear matter and discussions are presented in Sec.~\ref{sec:results}, and conclusions are provided in Sec.~\ref{sec:conclusions}.

\section{Sum rules of single-particle spectral functions}
\label{sec:sumrules}
Taking advantage of the analytical properties of the finite temperature one-body Green's function, one can write its Lehmann representation,
\begin{align}
g_k(\omega+ i \eta ) = \int_{-\infty}^{+\infty} \frac {d \omega'}{2 \pi } \frac {\mathcal{A}_k(\omega')}{\omega - \omega' + i \eta} \, ,
\label{eq:lehman}
\end{align}
in terms of the single-particle spectral functions, $\mathcal{A}_k(\omega)$ \cite{Abrikosov1965,Dickhoff08}. At the same time, 
the single-particle Green's function can be calculated, for any complex value $z$ of the energy, 
as the solution of  Dyson's equation 
 in terms of the self-energy $\Sigma_k$,
\begin{align}
g_k(z) = \frac {1}{z - \frac {\hbar^2 k^2}{2m} - \Sigma _k(z)} \, . 
\label{eq:dyson}
\end{align}

We decompose  $\mathcal{A}_k(\omega)$  in the sum of two positive functions, $\mathcal{A}^{<}_k(\omega)$ and 
$\mathcal{A}^{>}_k(\omega)$,
\begin{align}
\mathcal{A}_k(\omega) = \mathcal{A}^{<}_k(\omega) + \mathcal{A}^{>}_k(\omega) \, ,
\end{align}
defined like
\begin{eqnarray}
\mathcal{A}^{<}_k(\omega) = f(\omega) \mathcal{A}_k(\omega) \, , \nonumber \\
\mathcal{A}^{>}_k(\omega) = (1 - f(\omega)) \mathcal{A}_k(\omega) \, , 
\end{eqnarray}
where $f(\omega)$ is the Fermi-Dirac distribution,
\begin{align}
f(\omega) = \frac {1}{1+e^{\beta~(\omega-\mu)}}\, .
\end{align}
Here, $\mu$ is the chemical potential and $\beta=\frac{1}{T}$ is the inverse temperature.
The integration over all energies of $\mathcal{A}^{<}_k(\omega)$ provides the occupation probability of 
momentum $k$, 
\begin{align}
n_k = \int_{-\infty}^{+\infty} \frac {d \omega}{2 \pi } \mathcal{A}^{<}_k(\omega) \, . 
\end{align}
At zero temperature, $\mathcal{A}^{<}_k(\omega)$ [$\mathcal{A}^{>}_k(\omega)$] are directly linked to the hole (particle) spectral functions. 
 
The energy weighted sum rules for the single-particle spectral function can be derived by comparing  the asymptotic behavior of the real part of the Green's function at large $\omega$ obtained by two different methods. On the one hand, the Lehmann representation of Eq.~(\ref{eq:lehman}) becomes:
\begin{align}
\text{Re} g_k(\omega) = \frac {1}{\omega} & \left \{\int_{-\infty}^{+\infty}  \frac {d \omega'}{2 \pi} \mathcal{A}_k(\omega')  \right . \\
          &+ \left . \frac {1}{\omega} \int_{-\infty}^{+\infty} \frac {d \omega'}{2 \pi} \omega' \mathcal{A}_k(\omega') + ... \right \} \, .  \nonumber
\end{align}
On the other hand, the asymptotic expansion in terms of the self-energy in the Dyson equation yields
\begin{align}
\label{eq:realG}
{\rm Re} g_k(\omega) = \frac {1}{\omega} \left \{ 1 + \frac {1}{\omega} \left [ \frac {\hbar^2 k^2}{2m} 
+ \lim_{\omega \rightarrow \infty} {\rm Re} \Sigma_k( \omega ) \right ]+ ... \right \} \, .
\end{align}
Considering the coefficients of the asymptotic expansion order by order, one finds the corresponding sum rules associated to the moments of the spectral function, 
\begin{align}
m_k^{(n)} = \int_{-\infty}^{\infty} \frac {d \omega}{2 \pi} \omega^n \mathcal{A}_k(\omega) \, ,
\end{align}
at order $n$. Alternatively, mathematical expressions for higher-order moments can be obtained from the expectation value of nested commutators involving the Hamiltonian and creation and destruction operators \cite{Vogt2004,Duguet2012}.

The lowest order sum rule endows the spectral function with a probabilistic interpretation, 
\begin{align}
m^{(0)}_k = \int_{-\infty}^{\infty} \frac {d \omega}{2 \pi} \mathcal{A}_k(\omega) = 1 \, .
\end{align}
Physically, this sum rule indicates that all the single-particle strength is contained within the spectral function. Mathematically, it reflects the fact that $\mathcal{A}_k(\omega)$ is a positive-definite probability distribution for every momentum $k$. The first moment of the spectral function,
\begin{align}
m^{(1)}_k = \frac {\hbar^2 k^2}{2 m} + \lim_{\omega \rightarrow \infty } {\rm Re} \Sigma_k( \omega) \, ,
\label{eq:m1}
\end{align}
is directly related to the self-energy. In the Green's function formalism, the self-energy is usually decomposed in two pieces \cite{Dickhoff08},
\begin{align}
{\rm Re} \Sigma_k(\omega) = \Sigma^{\infty}_k
- P \int_{-\infty}^{+\infty}\frac{ d\lambda}{\pi} \frac {{\rm Im} \Sigma_k( \lambda    + i \eta)}{\omega - \lambda} \, .
\end{align}
The first term is an energy-independent contribution \cite{Polls1994}. The second term is a dispersive, energy-dependent contribution that can be computed from the imaginary part of $\Sigma$ alone. Here, and in the following, we consider only retarded self-energies. The corresponding time-ordered components can be obtained if needed by a well-known procedure \cite{Rios2009a}. Clearly, in the limit of very large energies, $\omega \rightarrow \pm \infty$, the second term decays like $1/\omega$ and $m^{(1)}_k$ is given in terms of the energy-independent contribution alone, 
\begin{align}
m^{(1)}_k = \frac {\hbar^2 k^2}{2 m} + \Sigma^{\infty}_k \, .
\label{eq:m1b}
\end{align}
The instantaneous part of the self-energy is analogous in structure to a Hartree-Fock term,
\begin{align}
\label{eq:genehf}
\Sigma^{\infty}_k &= 
\int \frac {d^3 k_1}{(2 \pi)^3} \langle \vec k \vec k_1 | V |  \vec k \vec k_1 \rangle_a ~ n_{k_1}   \\
                      &+ \frac {1}{2} \int \frac {d^3 k_1}{(2 \pi)^3} \frac {d^3 k_2}{(2 \pi)^3} 
                      \langle \vec k \vec k_1 \vec k_2
                        | W | \vec k \vec k_1 \vec k_2 \rangle_a n_{k_1} n_{k_2} \, ,\nonumber
\end{align} 
but is computed with a correlated momentum distribution $n_k$, which is affected by both correlations and temperature \cite{Polls1994}. We note that Eq.~(\ref{eq:genehf}) explicitly contains the effects of two- and three-body nuclear forces denoted by $V$ and $W$, respectively \cite{Duguet2012}. The subindex $a$ in the corresponding matrix elements indicates that they have been antisymmetrized. 

In nuclear physics, $m^{(0)}_k$ and $m^{(1)}_k$ have been explored extensively in a variety of contexts \cite{Polls1994,Frick2004,Rios2006,Duguet2012,Duguet2015}. In particular, $m^{(1)}_k$ is often called a ``centroid energy" and is closely related to effective single-particle energies in finite nuclei~\cite{Duguet2012}. In a sense, it provides single-particle energies which have been renormalized by single-particle fragmentation. In fact, if one takes the probabilistic character of the spectral function into account, $m^{(1)}_k$ is nothing but the mean of the distribution $\mathcal{A}_k(\omega)$. $m^{(1)}_k$ is also expected to be dependent on cut-off variation in the Hamiltonian \cite{Duguet2015}, as it is not an observable. 

Higher-order sum rules are obtained if more terms are retained in the expansion of Eq.~\eqref{eq:realG}. $m^{(2)}_k$ relates the spectral function to the imaginary part of the retarded self-energy:
\begin{align}
m^{(2)}_k &= \left [ \frac {\hbar^2 k^2}{2 m} + \Sigma^{\infty}_k \right]^2 
 - \int_{-\infty}^{+\infty}  \frac{d \omega}{\pi} \, {\rm Im}~\Sigma_k(\omega) \, ,
 \label{eq:m2}
\end{align}
where the dispersion relation for ${\rm Re} \Sigma_k( E)$ has been used \cite{Polls1994,Vogt2004}. The first term in the $m^{(2)}_k$ sum rule is simply $(m^{(1)}_k)^2$. In a probabilistic interpretation, the variance of the spectral function is related to the second moment
\begin{align}
\sigma_k^2 &=\int_{-\infty}^{+\infty} \frac{d \omega}{2\pi}  [ \omega - m^{(1)}_k ]^2   \mathcal{A}_k(\omega) 
= m^{(2)}_k - ( m^{(1)}_k )^2 \nonumber \\
&= - \int_{-\infty}^{+\infty}  \frac{d \omega}{\pi} \, {\rm Im}~\Sigma_k(\omega) \, .
\label{eq:sigma}
\end{align}
This result shows that the variance of $\mathcal{A}_k$ is nothing but the bulk integral of the imaginary part of the retarded self-energy \cite{Vogt2004}. In practical terms, the moments of the spectral function become increasingly difficult to compute numerically as the order increases, because they require accuracy in the high-energy tails. We shall see in the following that the $n=2$ moment and the variance are particularly sensitive to the short-range structure of nuclear forces. Along these lines, it is important to note that in the Hartree-Fock approximation $\text{Im }\Sigma=0$ and the variance is therefore null, $\sigma_k^2=0$. Consequently, a non-zero variance is already a clear signal of beyond mean-field correlations.

The sum rules discussed above, like spectral functions, are not observables in the strict sense of the word. As a consequence, we expect them to be sensitive to the underlying nuclear hamiltonian. 
This study is a proof-of-principle contribution that attempts at quantifying nuclear 
correlations as found in spectral function sum rules. The objective is to 
present properties which are only visible via analysis of the moments
of the spectral function, and which other many-body methods are 
possibly unaware of. This should in no way be considered as 
a mere ``selection procedure" for nuclear Hamiltonians, but rather as an analysis
of the imprints of nuclear correlations on spectral functions. 

\section{Results and discussions}
\label{sec:results}

\subsection{Details of the calculation}

All results reported in this paper have been obtained for symmetric nuclear matter at a
temperature of $T= 5$~MeV. Our formalism is based on a SCGF ladder resummation scheme that is formulated at finite temperature to avoid pairing instabilities \cite{Frick2003,RiosPhD}. 
The single-particle spectral functions in this approach are fully self-consistent, in the sense that they have been  used in the dressing of the two-particle propagator describing the intermediate states in the ladder effective interaction equation. The iterative procedure ensures self-consistency between the effective nucleon-nucleon interaction in the medium and the single-particle propagator. In all calculations of the in-medium $T-$matrix interaction, we have used partial wave decompositions  up to $J=4$ ($J=8$) in the dispersive (Hartree-Fock) contribution.

Our results have been computed at a single density of $\rho= 0.2$ fm$^{-3}$. This lies slightly above saturation and has been chosen to increase the effect of 3NFs on single-particle observables, which is otherwise relatively small \cite{Carbone2013a,Carbone2013b,Carbone2014}. This density is however low enough so that chiral forces are still applicable. All calculations with chiral forces are 
performed using the Entem-Machleidt (EM) N3LO 2NF as a starting point \cite{Entem2003,Machleidt2011}. 
In our approach, chiral 3NFs are taken into account as a density-dependent two-body force obtained by means of an average over the third particle. The average procedure, which takes care of the exchange effects, has been presented and discussed in detail in Refs.~\cite{Carbone2013a,Carbone2013b,Carbone2014}. 
The 3NFs are calculated at N2LO in the chiral expansion. We have used non-local regulators in three-body Jacobi coordinates set by a scale $\Lambda_\text{3NF}$ and an exponent $n$. Correlated momentum distributions have been used in the average procedure to obtain density-dependent one- and two-body effective forces. We have employed consistent 3NF low-energy constants $c_D$ and $c_E$. For results based on the bare EM N3LO 2NF and 3NFs, hereafter referred to as ``N3LO+3NF", the low energy constants $c_D=-0.201$ and $c_E=-0.614$ are extracted from the recent analysis of Ref.~\cite{Klos2016}. The 3NF cutoff is set at $\Lambda_\text{3NF}=500$ MeV and the regulator exponent is $n=3$. We note that our regulators are non-local as dictated by our present implementation of the 3N force and 
by the availability of corresponding low-energy constants. Developments to include 2N and 3N local forces like those described in Refs.~\cite{Navratil2007,Gezerlis2014,Epelbaum2015} are under way.

We will also present results obtained with SRG-evolved 2NFs complemented with 3NFs at the N2LO level. For these results, we choose the low-energy constants $c_D$ and $c_E$, the 3NF cut-off $\Lambda_\text{3NF}$, and the exponent $n$ from fits to the $^3$H binding energy and  $^4$He matter radius of Ref.~\cite{Hebeler2011}. In a way, this re-fit partially takes into account the effect of induced many-body forces by the SRG evolution. In particular, we will use as an example the calculations obtained with the EM N3LO 2NF evolved down to $\lambda=2$ fm$^{-1}$, and a N2LO 3NF with $\Lambda_{3NF}=2$ fm$^{-1}$ and $n=4$~\cite{Hebeler2011}. Hereafter, we refer to this combination as ``N3LO+SRG+3NF".

\begin{figure}[t!]
\begin{center}
\includegraphics[width=\linewidth]{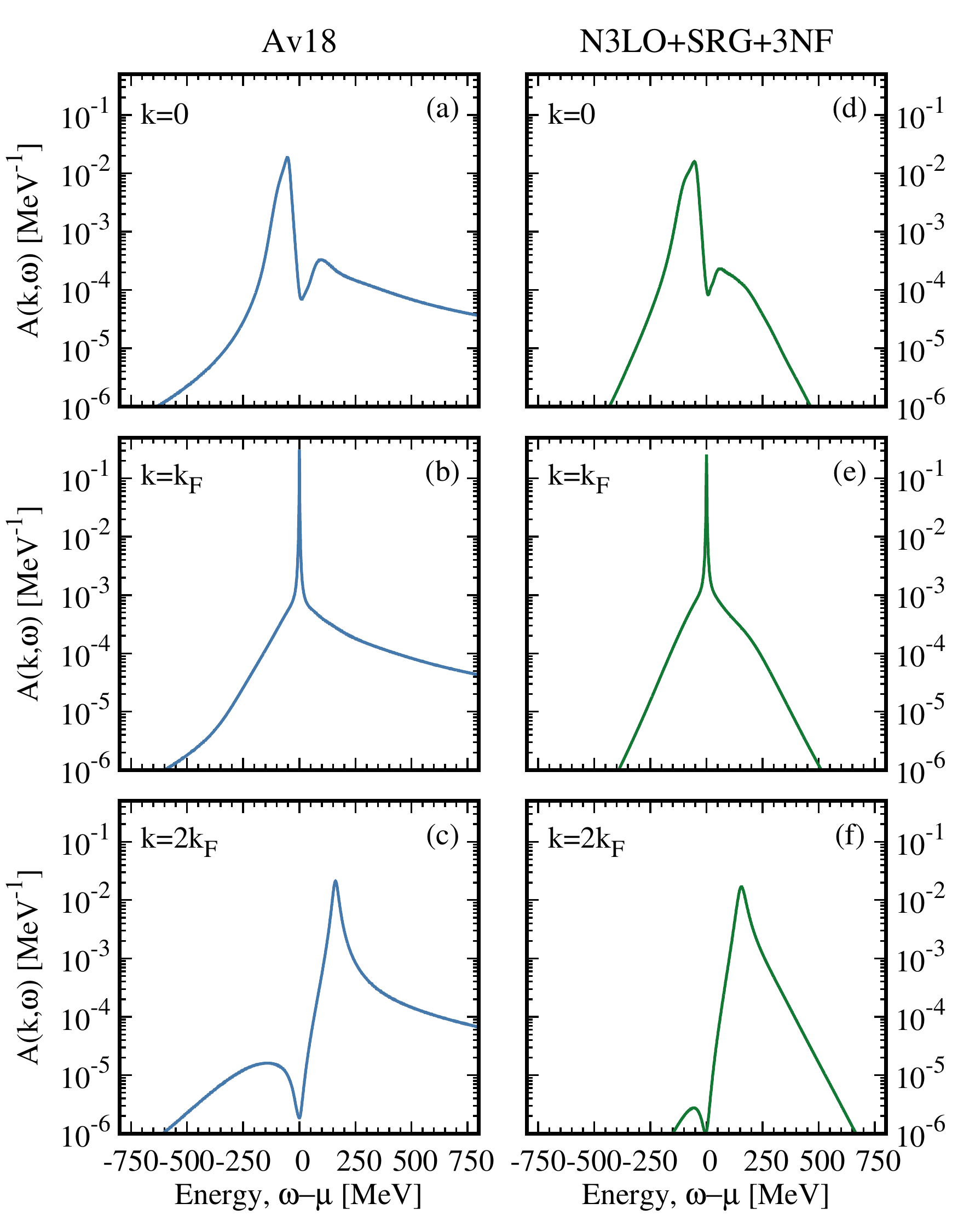}
\caption{\label{fig:spec} Single-particle spectral functions at $\rho=0.2$ fm$^{-3}$ and $T=5$ MeV for three different momenta. Left panels (a)-(c) correspond to the Av18 nucleon-nucleon interaction.  Right panels (d)-(f) show results for an SRG-evolved N3LO 2NF interaction down to $\lambda=2$ fm$^{-1}$ plus an N2LO 3NF. See text for further details.} 
\end{center}
\end{figure}

In the present discussion, we want to analyse the effect that different nuclear hamiltonians have on the moments of the spectral function. In this spirit, it is particularly useful to compare ``soft" and ``hard" interactions. On the one hand, we will use ``hard" phenomenological interactions, like CD-Bonn and Av18, which have already been used in previous studies of sum rules \cite{Polls1994,Frick2004,Rios2006}. On the other, we will use relatively soft chiral forces and use the SRG method to evolve them further to lower momentum scales. In a sense, we want to use the SRG as a handle for softness and identify any effects associated to the evolution towards lower scales in the sum rules. We will use in all cases the EM N3LO 2NF as a starting point for the SRG evolution. 

\subsection{Spectral functions}

We start our discussion by comparing, for three characteristic momenta, the single-particle spectral functions associated to two very different interactions. Figure~\ref{fig:spec} shows the single-particle spectral functions for $k= 0$ (top panels), $k=k_F$ (central panels) and $k=2 k_F$ (bottom panels) as a function of the energy variable $\omega-\mu$. 
We have selected two forces that would normally be chosen as examples of hard and soft interactions. Panels (a)-(c) correspond to the  hard Av18 interaction, whereas panels (d)-(f) are obtained from the SRG-evolved N3LO 2NF, complemented with a 3NF.

For all momenta, we find significant differences in the energy tails of the spectral functions, particularly at high positive excitation energies. Because N3LO+SRG+3NF is a very soft interaction, the high-energy tails die out well below $750$ MeV. In contrast, the high energy tails for Av18 are large, similar for the three-momenta considered and extend well into the GeV regime. At zero temperature, the tails at negative energies would correspond to the excitations of the $A-1$ system, $\mathcal{A}^<$. These are also more pronounced for Av18 than for N3LO+SRG+3NF. 
Unsurprisingly, we find that the fragmentation of single-particle states is very much suppressed for a soft hamiltonian as opposed to a hard force. We shall see in the following that the moments of the spectral function can provide a direct quantification of this fragmentation. 

The spectral functions at all momenta have a well-defined quasi-particle peak. For $k<k_F$, the peak  is well below the chemical potential, and the thermal function $f(\omega) \approx 1$. Therefore, the peak is provided by $\mathcal{A}^<_k(\omega)$. In contrast, well above the Fermi surface, similar arguments lead to the conclusion that the peak is caused by the $\mathcal{A}^>_k(\omega)$ component. $\mathcal{A}^<_k(\omega)$ and $\mathcal{A}^>_k(\omega)$ can be identified at zero-temperature as the hole and particle spectral function, respectively.  Whereas at zero temperature the $k=k_F$ spectral function would be a $\delta$ peak, the finite temperature $A_k(\omega)$ has a finite thermal width. Thermal effects are also responsible for the fact that, at $k \neq k_F$, the spectral functions are not zero at $\omega=\mu$. However, one can observe a well defined minimum for $\omega=\mu$, which is a reminder of the $T=0$ situation.

The self-consistency of the spectral functions, and the use of dispersion relations in the SCGF approach, guarantee that the sum rules are respected. In fact, the sum rule associated to the zeroth moment, $m^{(0)}_k=1$, is fulfilled to better than $0.1 \%$ in the complete momentum range, for all interactions used in the paper. This validates the accuracy of the numerical calculation and provides a useful test in terms of consistency.

\subsection{Kinetic energies}

\begin{table}[t!]
\begin{tabular}{r | c} 
Potential & Kinetic energy \\ 
& [MeV] \\ \hline
FFG & $27.1$ \\
N3LO+SRG &  $29.2$ \\
N3LO+SRG+3NF &  $29.4$ \\
N3LO & $36.4$ \\
N3LO+3NF & $36.7$ \\
CD-Bonn &  $38.9$ \\
Av18 & $47.1$ 
\end{tabular}
\caption{\label{table:kinetic} Kinetic energies per particle for the free Fermi gas as well as all the interactions considered in this work at $\rho=0.20$ fm$^{-3}$ and $T=5$ MeV. }
\end{table}

Many-body correlations induce large momentum components on the one-body momentum distribution, 
\begin{align}
n_k = \int_{-\infty}^{\infty} \frac {d \omega}{2 \pi} \mathcal{A}^<_k(\omega) \, .
\end{align}
In turn, these high momentum components enhance the kinetic energy, 
\begin{align}
\frac{K}{A} = \frac{1}{\rho} \int \frac {d^3 k}{(2 \pi)^3} \, \frac{k^2}{2m} n_k \, ,
\end{align}
which becomes larger than the corresponding free Fermi gas (FFG) value. The difference between the FFG kinetic energy and the correlated value is therefore an indicator of the ``hardness" or ``softness" of a nuclear hamiltonian \cite{Polls2000}. 

We report in Table~\ref{table:kinetic} the kinetic energies per particle at  $\rho=0.20$ fm$^{-3}$ and $T=5$ MeV for all the nuclear interactions of interest in this work. The table is sorted in increasing values of kinetic energy. In a sense, one goes from soft interactions at the top to hard interactions at the bottom. The FFG value quoted here takes into account the finite temperature, but is only $1.4$ MeV larger than the zero-temperature value. 

As expected, SRG-evolved interactions with and without three-body forces yield total kinetic energies which are comparable to the FFG. These forces have little strength in the high-momentum region and their corresponding kinetic energies are just about $\approx 2$ MeV larger than the FFG value. The calculations with an unrenormalized N3LO potential, in contrast, yield kinetic energies which are significantly higher, about $\approx 9$ MeV above the FFG prediction. This indicates the presence of more important high-momentum components. We note that in both the SRG-evolved and the unevolved cases the 3NF has a very small effect on the kinetic energy, increasing only by $\approx 0.2-0.3$ MeV. The high-momentum components in the CD-Bonn interaction yield a kinetic energy which is $\approx 11$ MeV above the FFG value. Finally, the kinetic energy of Av18 is $20$ MeV larger than the FFG. This large kinetic energy value is indicative of a significant departure from the uncorrelated momentum distribution. In the following, we discuss some of the imprints that these different high-momentum components have on the spectral function sum rules. 

\subsection{First moment}

\begin{figure}[t!]
\begin{center}
\includegraphics[width=\linewidth]{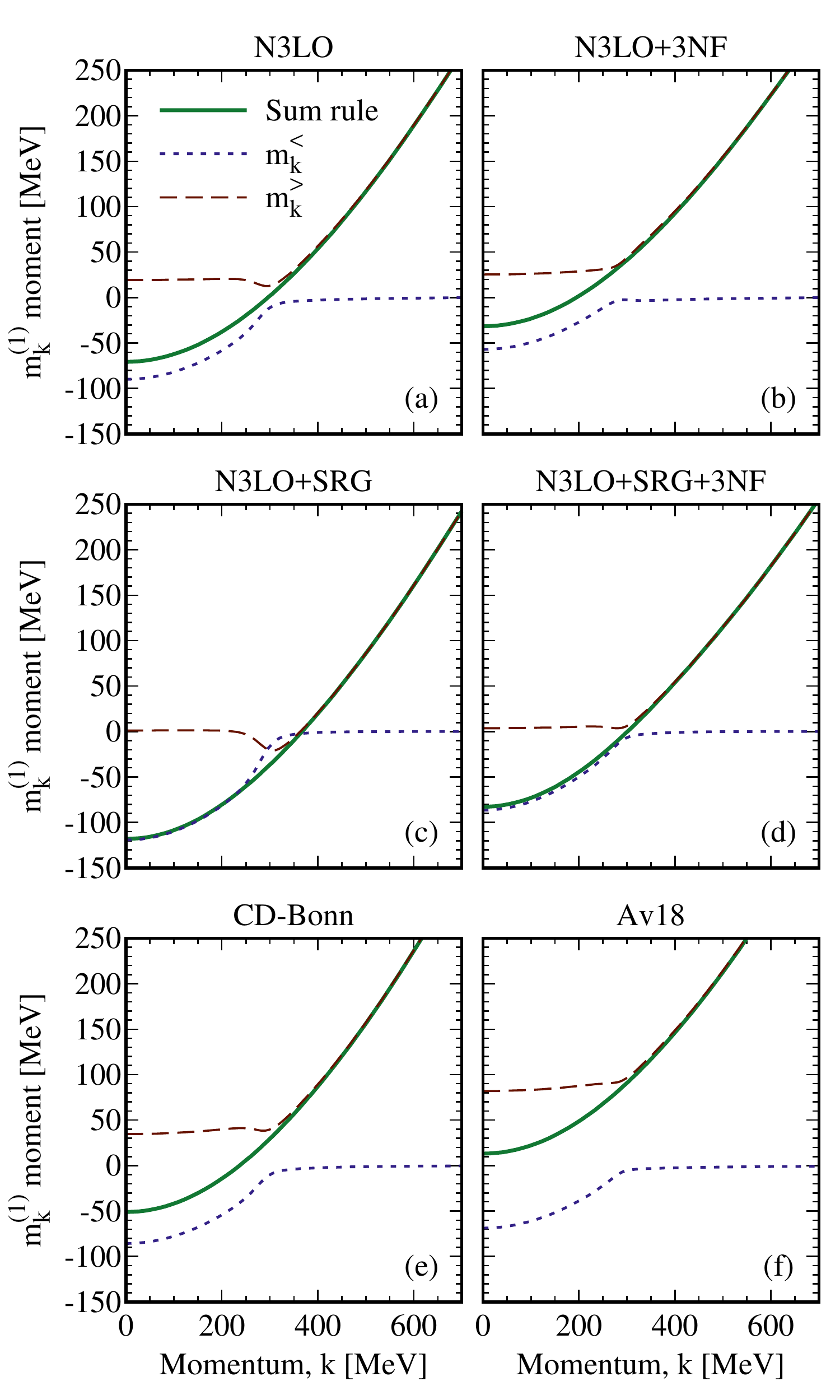}
\caption{\label{fig:m1}
Momentum dependence of the $m^{(1)}_k$ moment calculated at $\rho=0.2$ fm$^{-3}$ and $T=5$ MeV for the
(a) EM N3LO 2NF; (b) N3LO 2NF complemented with a 3NF; 
(c) SRG-evolved N3LO down to $\lambda=2$ fm$^{-1}$, 
(d) the same 2NF  complemented with a 3NF; 
(e) CD-Bonn and (f) Av18. 
In all panels, we show the first moment $m^{(1)}_k$ (solid line) and the right-hand side of Eq.~(\ref{eq:m1b}). The two sides of the sum rule are displayed, but they are indistinguishable. The $m^<_k$ (dot-dashed line) and $m^>_k$ (dashed line) contributions to $m^{(2)}_k$ are also shown. }
\end{center}
\end{figure}

Results for the $m^{(1)}_k$ sum rule are presented for different chiral interactions in panels (a)-(d) of Fig.~\ref{fig:m1}. Panels (e) and (f) show results for CD-Bonn and Av18, respectively. The $m^{(1)}_k$ sum rule is displayed as a function of momentum with a solid line. The results of the first moment of the spectral function and the right-hand-side of Eq.~(\ref{eq:m1b}) are plotted in the figure, but the curves lie one on top of each other and are indistinguishable. The sum rule is satisfied to better than $1\%$ for all momenta and interactions considered in the figure. For all forces, we find an increasing function of momentum due to the largely dominant kinetic term in Eq.~(\ref{eq:m1b}). 

The first moment is different depending on the Hamiltonian under consideration. This is to be expected on the grounds that $\Sigma_k^\infty$ is directly related to the nuclear force as dictated by Eq.~(\ref{eq:genehf}). We find an overall attractive contribution at $k=0$ for all forces, except for Av18. Calculations based on the hard Reid force \cite{Polls1994}, as well as previous calculations with Av18 at different densities \cite{Frick2004}, also yield positive $m^{(1)}_k$ sum rules. $\Sigma_k^\infty$ has a Hartree-Fock-like structure, which will necessarily yield repulsive results for hard-core forces. In accordance with this reasoning, we find that results associated to softer, SRG-renormalized, forces [panels (c) and (d)] induce more attractive $m^{(1)}_k$ moments than their unrenormalized counterparts [panels (a) and (b)]. Moreover, the effect of 3NFs is essentially repulsive. At $k=0$, we find that the 2NF-only N3LO [panel (a)] and N3LO+SRG [panel (c)] results yield $m^{(1)}_{k=0} \approx -70$ MeV and $\approx -118$ MeV, respectively. The corresponding 2NF+3NF results in panels (b) and (d) are shifted by about $\approx 35-40$ MeV to $m^{(1)}_{k=0} \approx -31$ MeV and $\approx -83$ MeV. We expect this shift to be density-dependent and sensitive to the 3NF structure. 

In analogy to previous studies, and in order to clarify the structure of the sum rule, we decompose the $m^{(1)}_k$ moment into two terms \cite{Polls1994,Frick2004}. The first one, $m^<_k$, is the first moment of $\mathcal{A}^{<}(\omega)$. The second, $m^>_k$, is the first moment due to $\mathcal{A}^{>}(\omega)$. In general, we find that $m^<_k$ is negative for low momentum and goes to zero above the Fermi momentum, $k_F =283$ MeV. This bodes well with the idea that $\mathcal{A}^{<}_k(\omega)$ carries the strength of the quasi-particle peak for momenta below $k_F$, whereas it is strongly suppressed at large $k$. 
On the contrary, $m^>_k$  is generally positive below $k_F$, due to the tail of $\mathcal{A}^{>}_k(\omega)$ at high positive energies, $\omega > \mu$. The flatness of $m^>_k$ below the Fermi momentum indicates that this high-energy tail is rather momentum independent. Around $k=k_F$,  $m^>_k$ can dip down and even become negative, particularly for soft interactions with less prominent high-energy tails. For $k > k_F$, $m^>_k$   grows steadily because the sum rule integral now captures the quasi-particle peak.
 
While the qualitative momentum dependence of $m^>_k$ is independent of the Hamiltonian, we find that the size of this contribution changes substantially from force to force. In particular, there is a substantial dependence on the renormalization scheme used to tame the 2NF. We focus first on the unrenormalized results based on the N3LO 2NF of panel (a). The $m^>_k$ contribution for $k < k_F$ is, as mentioned earlier, momentum independent and close to $m^>_k \approx 20$ MeV. The results of panel (b), where $m^>_k \approx 25$ MeV below the Fermi surface, suggest that 3NFs do not substantially alter the high-energy tails, in accordance with the results of Refs.~\cite{Carbone2013b,CarbonePhD}.

When the SRG is used to renormalize the N3LO interaction (with an SRG cutoff of $2$ fm$^{-1}$), the interaction becomes softer and short-range correlations are tamed. This is reflected in much shorter high-energy tails in the spectral function, as discussed in the context of Fig.~\ref{fig:spec}. In consequence, the $m_k^>$ contributions in panel (c) are heavily suppressed, and one finds that $m^>_{k<k_F} \approx 0$. This result does not change when 3NFs are included [panel (d)]. In contrast, the hard interactions of panels (e) and (f) have extended positive energy tails, and $m^>_{k<k_F}$ is relatively large. For CD-Bonn, for instance, one finds $m^>_{k<k_F} \approx 35$ MeV, whereas Av18 yields a much larger contribution, $m^>_{k<k_F} \approx 82$ MeV.

All in all, our results indicate that the first moment of the spectral function is sensitive to the hard- or soft-core nature of nuclear forces. Negative $m_k^{(1)}$ at momenta below $k_F$ are indicative of attractive, soft interactions. Since the $m^<_k$ contribution is always attractive for $k<k_F$, the main driver for the sign of  $m_k^{(1)}$ below the Fermi surface is the size of the $m^>_{k<k_F}$ term. In other words, the strength of the $\mathcal{A}^>$ component of the spectral function for $k<k_F$ is a good proxy for the softness of the interaction. For the density that we have chosen, soft chiral interaction have $m^>_{k<k_F} \lesssim 20$ MeV. In contrast, hard phenomenological interactions have $m^>_{k<k_F} \gtrsim 40$ MeV. Furthermore, the inclusion of 3NFs mostly yields more repulsive quasi-particle peak energies, and affects only mildly the high-energy tails in the spectral function. 

\subsection{Second moment}

\begin{figure}[t!]
\begin{center}
\includegraphics[width=\linewidth]{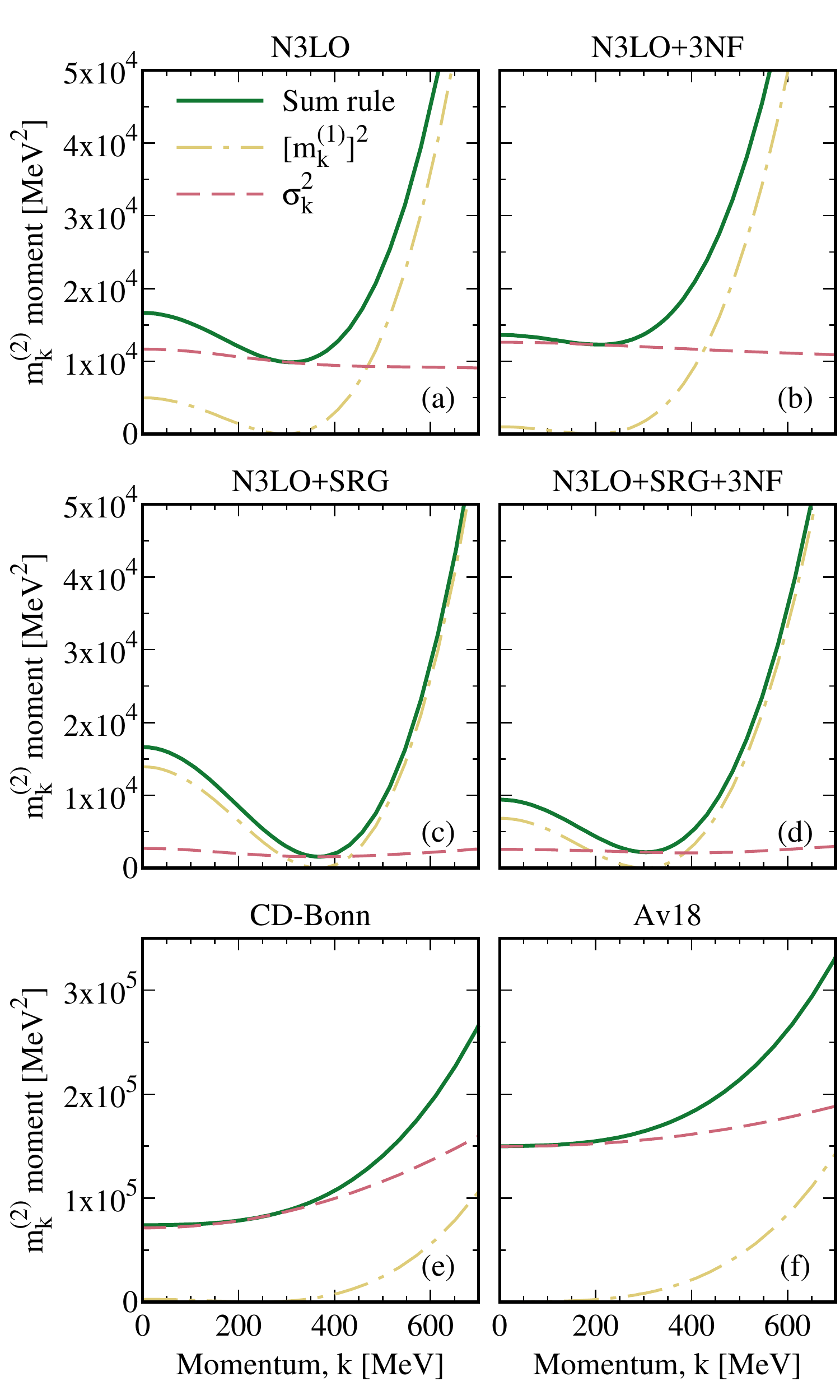}
\caption{\label{fig:m2}
Momentum dependence of the $m^{(2)}_k$ moment calculated at $\rho=0.2$ fm$^{-3}$ and $T=5$ MeV for the same interactions as Fig.~\ref{fig:m1}. 
In all panels, we show the first moment $m^{(2)}_k$ (solid line) and the right-hand side of Eq.~(\ref{eq:m2}). The two sides of the sum rule are displayed, but they are indistinguishable. The $(m^{(1)}_k)^2$ and the $\sigma_k^2$ contributions to $m^{(2)}_k$ are also shown. }
\end{center}
\end{figure}

We show $m^{(2)}_k$ as a function of momentum in Fig.~\ref{fig:m2}. Note the difference in vertical scale between panels (a)-(d) and (e) and (f). In accordance to the results of the lower order moments, the sum rule for $m_k^{(2)}$ is respected very well for all interactions and momenta considered here. Both the moment and the right-hand side of Eq.~(\ref{eq:m2}) are displayed in the figure, but they are indistinguishable. 

As the integral of the product of two positive definite factors, $m^{(2)}_k$ is always positive. As shown in Eqs.~(\ref{eq:m2}) and (\ref{eq:sigma}), the second moment can be split into two physically motivated terms. The first term is the square of the first moment, $[m^{(1)}_k]^2$, which equals the generalized Hartree--Fock contribution to the self-energy. We show this term in dash-dotted lines in Fig.~\ref{fig:m2}. The second contribution is the variance of the spectral function, $\sigma_k^2$, which is displayed in dashed lines. 

We discuss first the overall magnitude and momentum dependence of the second moment. This is markedly different for traditional phase-shift equivalent potentials like Av18 or CD-Bonn [panels (e) and (f)] than for chiral N3LO interactions, independently of whether they have been renormalized or not [panels (a)-(d)]. For chiral forces, $m^{(2)}_k$ is a decreasing function of momentum at low momentum. Close to $k \approx 0$, $m^{(2)}_k \approx (1-2) \times 10^4$  MeV$^2$, for all chiral forces. The repulsive effect of  3NFs [panels (b) and (d)] is reflected in a smaller value of $m^{(2)}_k$ at low momenta in comparison with 2NF-only results [panels (a) and (c)]. 

Around $k \approx 300$ MeV, a clear minimum develops for all chiral interactions. The value of $m^{(2)}_\text{min}$ is very sensitive to the renormalization procedure. Unrenormalized interactions yield $m^{(2)}_\text{min} \approx 10^4$ MeV$^2$, whereas SRG renormalized interactions have much smaller minima $\approx 10^3$ MeV$^2$. For momenta above the Fermi surface, the second moment increases steeply, following the dominant momentum dependence of $[m^{(1)}_k]^2$. 

The second moment for traditional forces has a very different structure. No sharp minimum develops as a function of momentum. In contrast to the values observed for chiral forces, below the Fermi surface we find large values, $m^{(2)}_k \approx (0.8-1.5) \times 10^5$ MeV$^{2}$ for CD-Bonn and Av18. Further, the second moment is a steeply increasing function of momentum. This is due largely to the monotonically increasing momentum dependence of $\sigma_k^2$. 

The difference between chirally motivated forces and hard core interactions can be explained in terms of the relative contributions of the first moment and the variance of the spectral function. On the one hand, we have a contribution to $m^{(2)}_k$ that is precisely the square of $m^{(1)}_k$. As long as $\sigma_k^2$ is small, the location of the zero of the $m^{(1)}_k$ (if there is one) coincides with the minimum of  $m^{(2)}_k$. We note that the position of this zero is sensitive to the Fermi momentum, but also to the structure of the potential, as can be understood by looking at the right-hand side of Eq. (\ref{eq:m1b}). Figure~\ref{fig:m1} clearly illustrates that traditional forces with strong short-range potentials can have a zero in $m_k^{(1)}$ which are shifted from $k_F$ (or even have no zeros at all, like Av18).

On the other hand, the variance of the spectral functions for soft chiral interactions is very different than that of traditional potentials. First, the variance is smaller for softer forces than it is for harder ones. In fact, there is a clear hierarchy between the different types of interactions considered in this paper. We have already noted that, close to $k=0$, traditional forces have values $\sigma_k^2 \approx 10^5$ MeV$^2$. N3LO and N3LO+3NF, in contrast, have values which are about an order of magnitude smaller, $\sigma_k^2 \approx 10^4$ MeV$^2$. Further, SRG-evolved interactions have even lower variances, $\sigma_k^2 \approx 10^3$ MeV$^2$. In terms of momentum dependence, hard forces have increasing variances as a function of momentum, whereas chiral forces have much more constant values. 

\begin{figure}[t!]
\begin{center}
\includegraphics[width=0.9\linewidth]{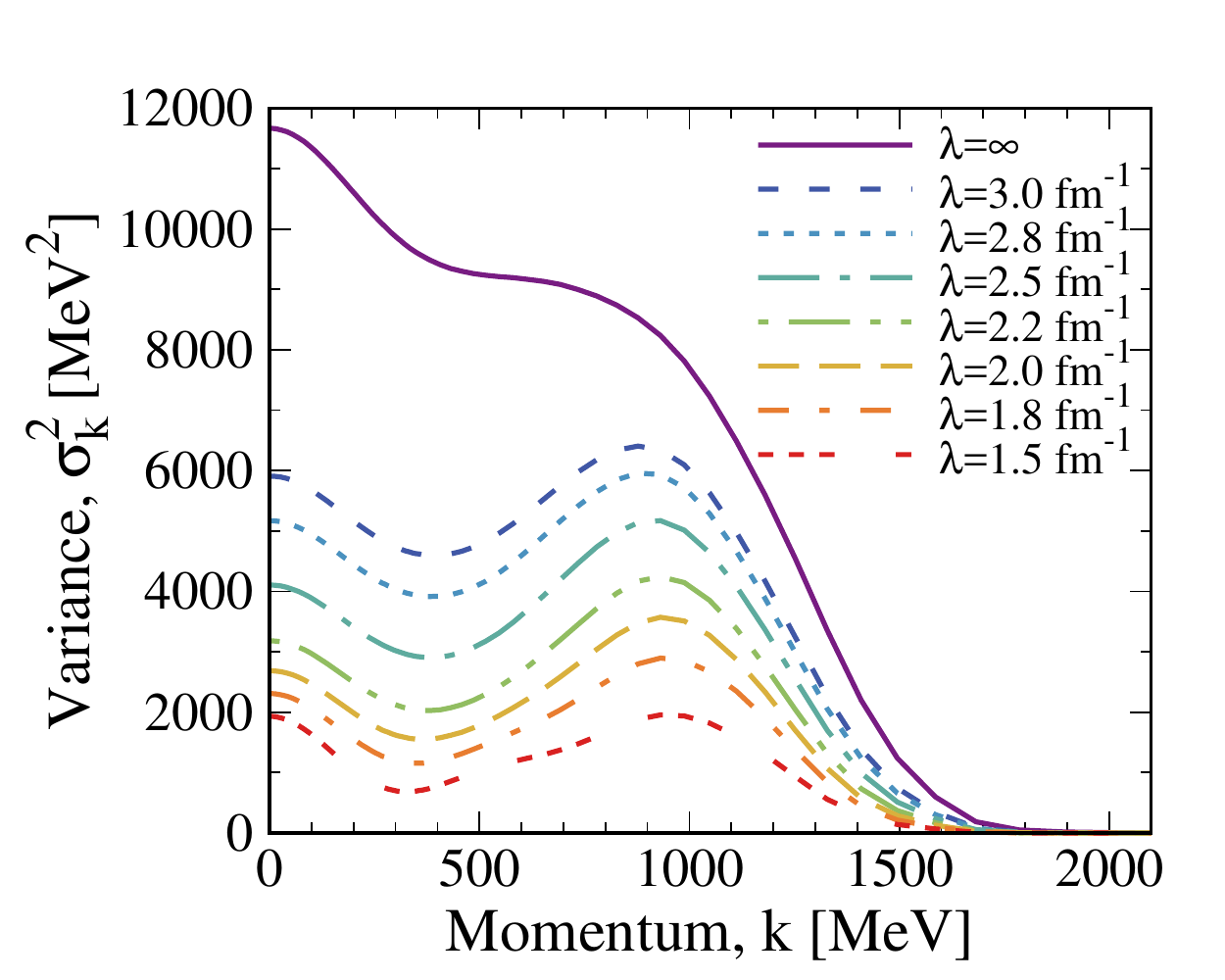}
\caption{\label{fig:variance} Momentum dependence of the variance, $\sigma_k^2$, of the 
spectral function $\mathcal{A}_k(\omega)$ at $\rho=0.2$ fm$^{-3}$ and $T=5$ MeV. Results are displayed for 2NF-only calculations with the bare N3LO force (solid lines) and SRG-evolved N3LO forces in the range $\lambda=3$ fm$^{-1}$ to $\lambda=1.5$ fm$^{-1}$. } 
\end{center}
\end{figure}

We can gain further insight on how this hierarchy develops by considering the variance of $\mathcal{A}_k(\omega)$ obtained with a variety of SRG-evolved interactions. In Fig.~\ref{fig:variance}, we plot $\sigma_k^2$ as a function of momentum for N3LO 2NFs (corresponding to the solid line, $\lambda=\infty$), and for SRG-evolved interactions in scales ranging from $\lambda=3$ fm$^{-1}$ to $\lambda=1.5$ fm$^{-1}$. We find that there is a clear one-to-one correspondence between $\lambda$, the renormalization scale cut-off, and the value of  the variance. Smaller scales yield smaller values of $\sigma^2_k$, in an almost linear correspondence. Further, we also note that all the variances go to zero for a momentum of $k \approx 1.5$ GeV. This is in contrast to the behavior observed for hard forces, where the variance increases steadily as a function of momentum.

This result is relevant beyond the present context of nuclear matter calculations. First, because in a mean-field picture the variance is zero, one expects  $\sigma_k^2$ to be directly related to beyond mean-field correlations, independently of the system under study. Figure~\ref{fig:variance} indicates that this is indeed the case. The variance is proportional to the renormalization of short-range correlations, at least in the case where the SRG is used. 
Further, the variance of the spectral function is a telling sign for the fragmentation of single-particle states. In other words, very fragmented states will have large $\sigma_k^2$, as the contributions away from the quasi-particle peak contribute with $(\omega-[m^{(1)}_k])^2$. In contrast, very narrow, quasi-particle-like states will have small variances. As seen in the context of Fig.~\ref{fig:spec}, chiral forces yield spectral functions with far lower tails and a stronger quasi-particle nature. By further renormalizing the short-range structure of these interactions, the spectral functions become increasingly more peaked and the variance further decreases. These arguments are independent of the many-body system, so one would expect similar results in finite nucleus calculations \cite{Duguet2012,Duguet2015}, at least within the size of the active model space used in the calculations. 

In addition to the difference in absolute magnitude, $\sigma^2_k$ is rather different in terms of momentum dependence when one compares hard forces to chirally motivated ones. The latter are regularised at large momenta ($\Lambda=500$ MeV for the EM force). It is therefore natural to expect that the spectral function lacks support for momenta well beyond $\Lambda$. The results of Fig.~\ref{fig:variance} indicate that the width of the spectral function is maximum at the origin, then again around $k \approx 1$ GeV and finally dies out beyond $1.5$ GeV. Rather interestingly, this momentum dependence is relatively independent of the SRG evolution scale. 

\begin{figure}[t!]
\begin{center}
\includegraphics[width=0.9\linewidth]{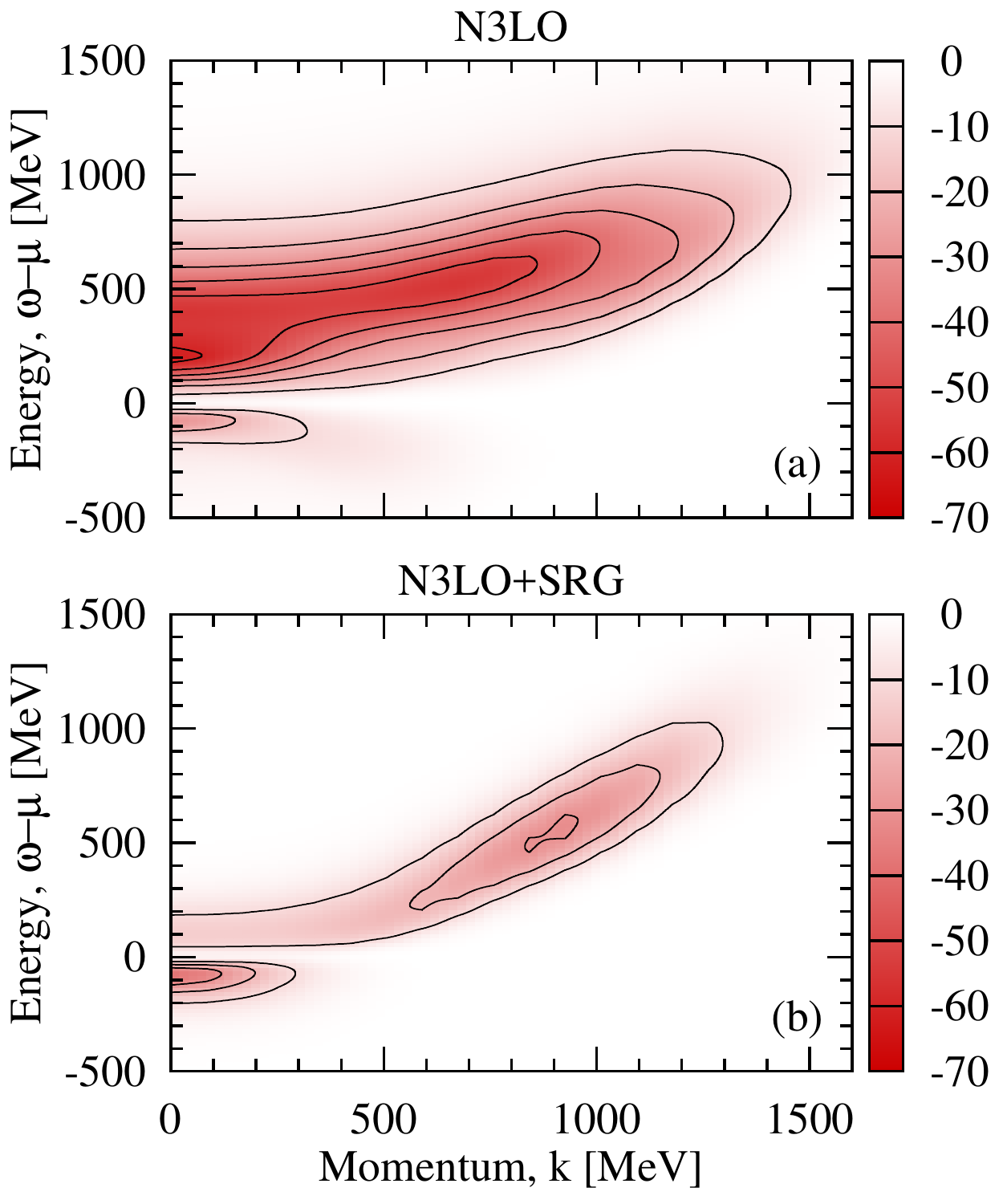}
\caption{\label{fig:self} Contour density plot of the momentum and energy dependence of $\text{Im }\Sigma_k(\omega)$ at  $\rho=0.2$ fm$^{-3}$ and $T=5$ MeV. Panel (a) corresponds to the N3LO interaction, whereas panel (b) results are obtained with an SRG-evolved N3LO 2NF with $\lambda=2$~fm$^{-1}$. Contour levels are spaced by $10$ MeV. }
\end{center}
\end{figure}

Further insight into the properties of the second moment and the variance of the spectral function can be gained by looking at $\text{Im }\Sigma_k$. Equation~(\ref{eq:sigma}) indicates that, for a fixed $k$, the variance is equal to the energy integral of the $\text{Im }\Sigma_k$. Figure~\ref{fig:self} shows a contour density plot of ${\rm Im} \Sigma_k(\omega)$ as a function of momentum and energy for N3LO 2NF [panel (a)] and for an SRG-evolved N3LO 2NF with a renormalization scale of $\lambda=2$ fm$^{-1}$. The color codes of the two panels are the same, and the contours are spaced in units of $10$ MeV. The support of the two self-energies is limited to a relatively narrow range of energies and momenta. For $\omega < \mu$, the hole contributions to $\text{Im }\Sigma_k$ becomes zero below $\omega - \mu \approx -500$ MeV. The particle part ($\omega>\mu$) of the self-energy extends to $1.5$ GeV at most. Similarly, and in accordance to our previous discussion regarding Fig.~\ref{fig:variance}, the self-energy is basically zero for momenta $k > 1.5$ GeV. This is unsurprising in terms of the tamed short-range structure of these potentials which, together with the phase space dictated by density, sets the scale for $\text{Im }\Sigma_k$ at large energies \cite{FrickPhD}. For reference, $\text{ Im} \Sigma_k$ for hard-core potentials like Av18 and CD-Bonn would typically extend to $\omega \approx 10^4$ MeV.  

The SRG evolution has a rather dramatic effect on the energy extent and the overall size of the particle component of $\text{Im }\Sigma_k$. N3LO has a deeper and more extended $\text{Im }\Sigma_k$, reaching minima close to $\approx- 70$ MeV around $\omega - \mu \approx 400$ MeV. In contrast, the SRG-evolved results provide a much shallower self-energy, with $\text{Im }\Sigma_k > -40 $ MeV throughout. Further, the support for $\text{Im }\Sigma_k$ in the energy-momentum plane is narrower, and concentrated on a diagonal band with a width of less than $200$ MeV. This band is in correspondence with the quasi-particle peak, which in this case plays a dominant role in the spectral function. These results are consistent with the behaviour of the variance in Fig.~\ref{fig:variance}: the integral of a narrow and shallow $\text{Im }\Sigma_k$ will yield small variances. In contrast, a relatively extended self-energy in energy space will provide larger $\sigma_k^2$. In a sense, the variance of the spectral functions quantifies the extent and structure in energy of $\text{Im }\Sigma_k$.

\subsection{Similarity renormalization group and three-nucleon forces}

The spectral function moments are not observables in a quantum mechanical sense~\cite{Duguet2012,Duguet2015}. In principle, they should depend (and possibly be sensitive to) the renormalization scale of the hamiltonian. For true observables, the dependence with the SRG can be taken as an indication of the theoretical uncertainty in the calculation. In the context of finite nuclei and nuclear matter, it has been argued that the scale dependence of observables due to the hamiltonian should be reduced if 3NF, including those induced in the renormalization process of the 2NF, are consistently included  \cite{Hebeler2013,Jurgenson2011}. For non-observables, one does not necessarily expect an independence on the SRG scale. In particular, for the first moment, finite nuclei calculations indicate that the scale-dependence is still relevant \cite{Duguet2015}. One might however still look at this scale dependence, with and without 3NF, to identify potential trends and quantify how single-particle fragmentation evolves with renormalization.

\begin{figure}[t!]
\begin{center}
\includegraphics[width=\linewidth]{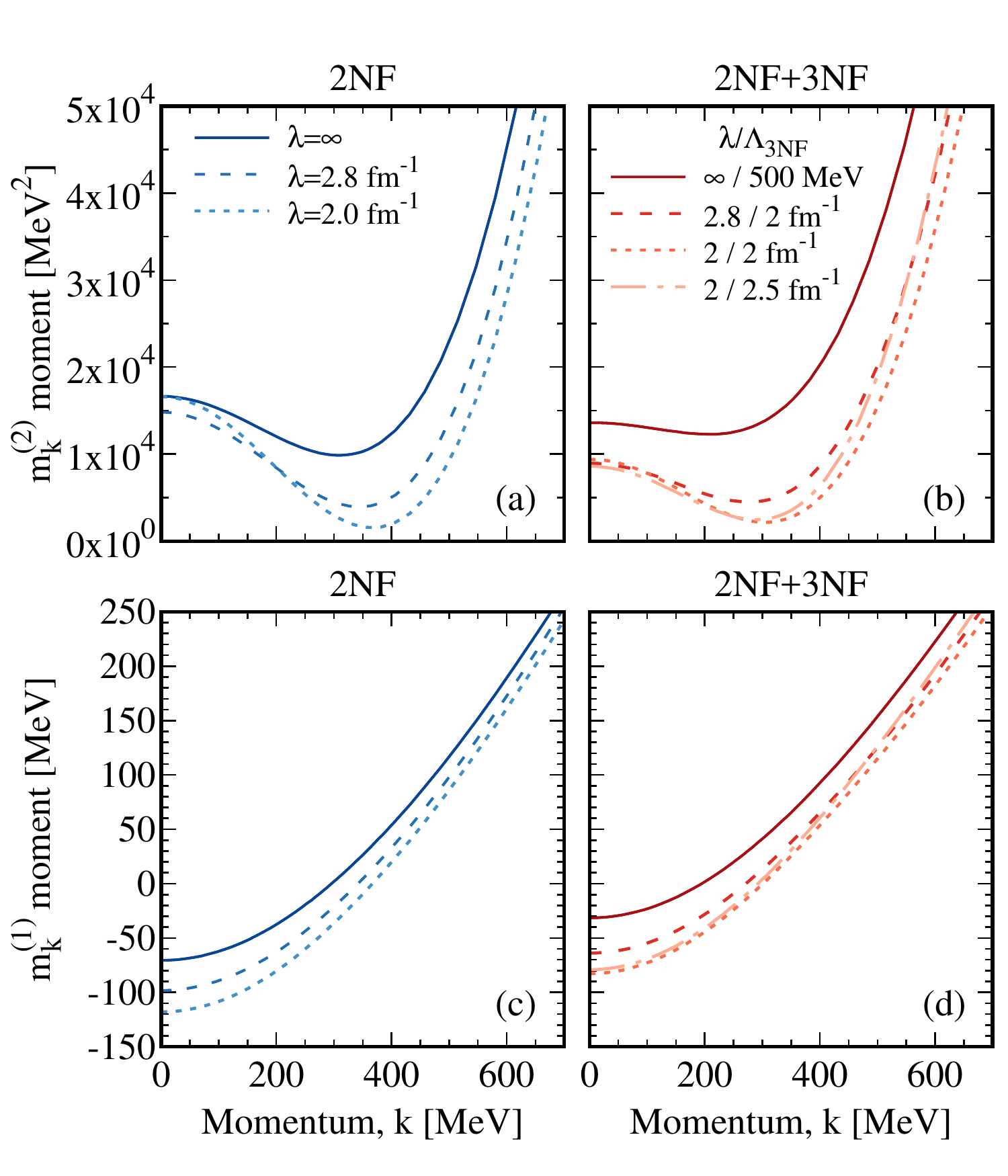}
\caption{\label{fig:m12_srg}
Momentum dependence of the $m^{(2)}_k$ [panels (a) and (b)] and $m^{(1)}_k$ [panels (c) and (d)] moment calculated at $\rho=0.2$ fm$^{-3}$ and $T=5$ MeV for several interactions. Panels (a) and (c) show results for non-evolved and  SRG-evolved EM N3LO interactions to a scale $\lambda$. Panels (b) and (d) show results of calculations with non-evolved and with SRG-evolved 2NFs as well as refit 3NFs. In these panels, $\lambda$ denotes the SRG scale and $\Lambda_\text{3NF}$ the scale of the 3NF regulator. See text for details. }
\end{center}
\end{figure}

Unlike finite nuclei calculations, here we do not consider the effect of induced 3NFs. We explore the possibility that the renormalization induced by refit 3NFs already restricts the scale dependence of the moments. To this end, we present in panels (a) and (c) of Fig.~\ref{fig:m12_srg} the $m^{(2)}_k$ and $m^{(1)}_k$ moments for 2NF-only calculations for the bare EM 2NF (solid lines) and for two different SRG scales, $\lambda=2.8$ fm$^{-1}$ (dashed line) and  $\lambda=2$ fm$^{-1}$ (dotted line). Panels (b) and (d) contain results which include the effect of 3NFs. We show results for the bare N3LO and 3NF  (solid lines), which correspond to the ``N3LO+3NF" results presented earlier with $\Lambda_\text{3NF}=500$ MeV and $n=3$ in the regulator function. We then present three curves with different cut-off scales. First, we run the SRG on the 2NF down to $\lambda=2.8$ fm$^{-1}$ and produce the dashed lines in panels (b) and (d) with a 3NF with $\Lambda_\text{3NF}=2$ fm$^{-1}$. Second, we show a case with $\lambda=2$ fm$^{-1}$ and $\Lambda_\text{3NF}=2$ fm$^{-1}$ (dotted line). The third case has the same 2NF cut-off, $\lambda=2$ fm$^{-1}$, but a different 3NF cut-off, $\Lambda_\text{3NF}=2.5$ fm$^{-1}$ (dash-dotted line). The low-energy constants $c_D$ and $c_E$ and the exponent of the regulator, $n=4$, for all the SRG-evolved cases are chosen following Ref.~\cite{Hebeler2011}. As mentioned earlier, in all cases 3NFs have  been included using a correlated momentum distribution in the averaging procedure and an internal momentum-dependent regulator \cite{Carbone2014}. 

Panels (a) and (c), based on 2NF-only results, are in line with our previous comments. The SRG running of the 2NF yields softer and softer forces, which in turn induce more and more attraction in the $m^{(1)}_k$ moment at all momenta [panel (c)]. We find that the SRG generally reduces the second moment [panel (a)] at large momenta. At small momenta, $m^{(2)}_k$ first decreases ($\lambda=2.8$ fm$^{-1}$) and then increases ($\lambda=2$ fm$^{-1}$), contrary to the value
of the cut-off scale. Since this cannot be due to $\sigma^2_k$ [see Fig.~\ref{fig:variance}], we attribute the effect to the $[m^{(1)}_k]^2$ term, which grows with $\lambda$ due to the increasing attractive $m^{(1)}_k$ contribution. 

For the 2NF+3NF results in panel (d), we find a similar trend. Decreasing the SRG scale, we find more attractive $m^{(1)}_k$ contributions. The dependence in $\lambda$ is relatively similar, particularly in the low-momentum region ($k<200$ MeV). In going from $\lambda=2.8$ fm$^{-1}$ to $\lambda=2$ fm$^{-1}$, $m^{(1)}_0$ decreases by $\approx 18.4$ MeV. This is to be compared to the decrease of about $\approx 19.7$ MeV in the 2NF-only case. In contrast, $m^{(2)}_0$ increases by $1.8 \times 10^3$ MeV$^2$ with 2NF only, but the increase when 3NF are included reduces to $\approx 8 \times 10^2$ MeV$^2$. These differences are not constant as a function of momenta. Qualitatively, the differences in $m^{(2)}_k$ for different SRG scales are similar independently of whether 3NFs are included or not. We note that the dependence on $\Lambda_\text{3NF}$ is also remarkably small and confined to the large-momentum region. 

Overall, we find a similar scale dependence with and without 3NFs, in a restricted range of SRG cut-offs. This points to the importance of many-body forces in the moments of the spectral function in infinite matter. Along these lines, the difference between the unrenormalized and the SRG-evolved results are a potential indication of the size of induced 3NF effects and, possibly, of truncations in the many-body scheme. 

\subsection{Running sum rules}

\begin{figure}[t!]
\begin{center}
\includegraphics[width=\linewidth]{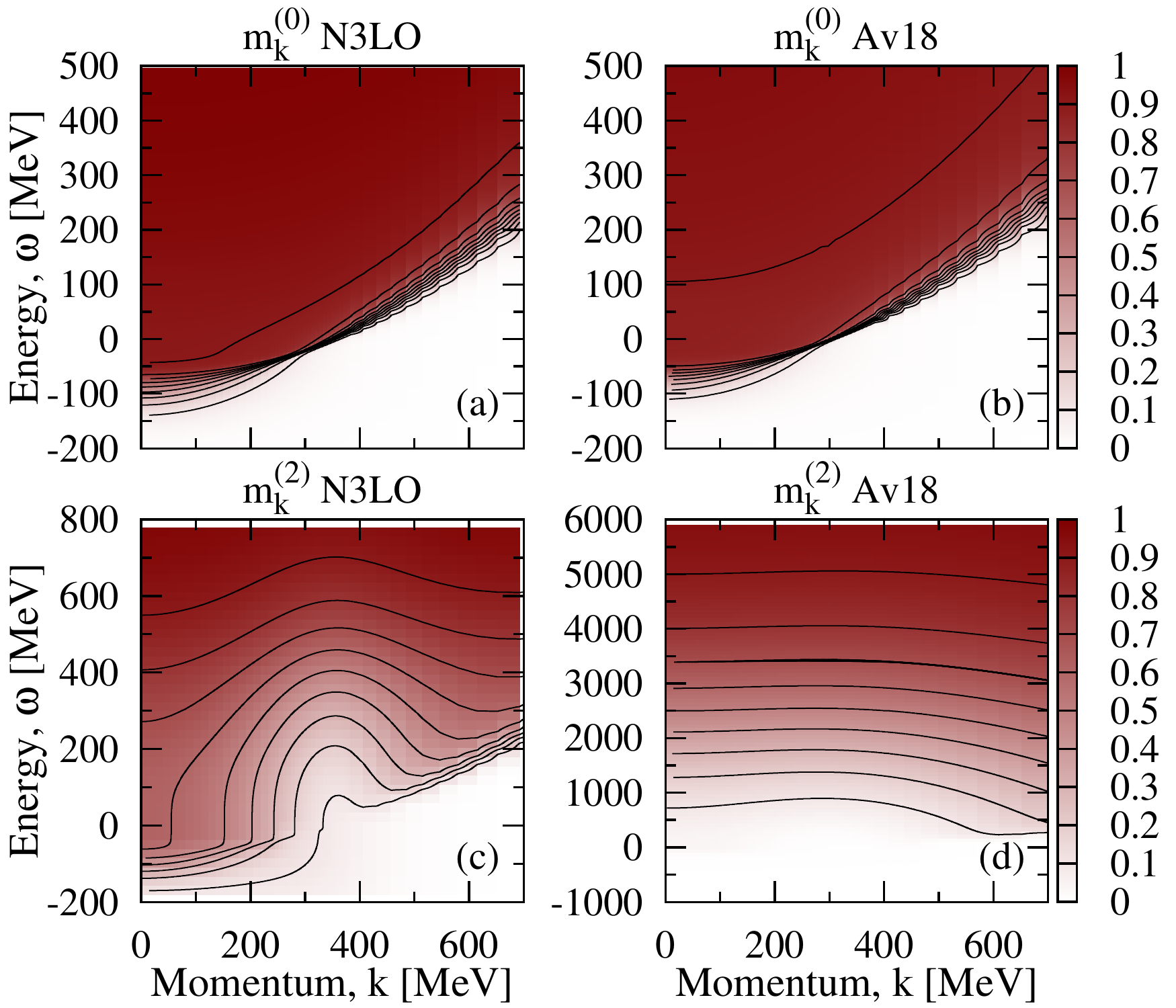}
\caption{\label{fig:runsum} 
Top panels: saturation of the $m^{(0)}_k$ sum rule as a function of momentum and upper limit of the integration for the (a) N3LO and (b) Av18 2NFs. 
Bottom panels: saturation of the $m^{(2)}_k$ sum rule for the (c) N3LO and (d) Av18 2NFs. 
} 
\end{center}
\end{figure}

Finally, it is also useful to look at how the sum rules saturate with increasing energy. For the $m^{(0)}_k$ and $m^{(1)}_k$ moments, the running sum rules have been studied extensively in Refs.~\cite{Polls1994,Frick2004,Rios2006}. With the hard interactions used in those references, an integration of up to $\approx 0.5$ ($5$) GeV is needed to saturate $m^{(0)}_k$ ($m^{(1)}_k$) to within $90 \, \%$. Naively, one expects that even higher energies will be needed to saturate the second moment. We explore the saturation of $m^{(0)}_k$ and $m^{(2)}_k$ in Fig.~\ref{fig:runsum}. Each panel shows a contour density plot of the exhausted sum rule as a function of momentum for each energy. The results have been normalized to the total value of the sum rule, and each contour line represents a $10 \, \%$ increase towards the $100 \, \%$ saturated value.  Panels (a) and (c) show the N3LO 2NF results for the zeroth and second moments, whereas panels (b) and (d) display the same data for Av18. 

The N3LO results for the two sum rules show that these are exhausted immediately after the quasi-particle peak is integrated. This is particularly true for $m^{(0)}_k$, where the true value of the sum rule is achieved within $\approx 150$ MeV of the peak. For $m^{(2)}_k$, one has to integrate $400-500$ MeV away from the peak to saturate the sum rule. There is a clear momentum dependence of the results, associated to the evolution of the quasi-particle peak with momentum. Further, for $m^{(2)}_k$, the region around $k=k_F$ requires larger energies to saturate the sum rule. This effect is due to the narrowing of the quasi-particle in this region, which in turn translates into a build-up of flat high-energy tails that require increasing energies. In any case, it is clear that the prominence of the quasi-particle peak in the N3LO results substantially dominates the sum rule saturation profile. Energies below $1$ GeV suffice to get over $90 \, \%$ of the strength for both sum rules, in contrast to previously found results with harder forces. With SRG-renormalized interactions, even lower integration limits are required. 

This situation is in contrast to the typical results of a hard interaction like Av18 [panels (b)-(d)]. Note the difference in vertical scales in the bottom panels.
For $m^{(0)}_k$ [panel (b)], we find that the sum rule saturates more slowly for this hard-core interaction. Whereas $\approx 80 \, \%$ of the moment is integrated within $50$ MeV of the peak, a saturation of $90 \, \%$ requires much higher energies than the corresponding N3LO counterpart. One needs to integrate at least beyond $150$ MeV above the quasi-particle peak to saturate. The situation is even more extreme for the second moment. To get to a $90 \, \%$ value, one needs to integrate at least up to $5$ GeV (in contrast to the $<1$ GeV needed for N3LO). Below the Fermi surface, $k<k_F$, we find that the same amount of energy is needed to exhaust the sum rules to a given level, independently of momentum. This indicates that the contribution of the quasi-particle peak, which is strongly momentum-dependent, is less relevant for the second moment. Above the Fermi surface, we find that less energy is required to saturate the second moment. At these momenta, the peak is at positive energies and its width increases with momentum. In consequence, high-energy tails are expected to decrease and contribute less to the second moment. We note that similar results are found for other hard interactions, like CD-Bonn. We have also analysed the running $m_k^{(1)}$ sum rule, which presents a more complex structure due to the change in sign.

\section{Conclusions}
\label{sec:conclusions}

We have analyzed the behavior of the energy weighted sum rules of single-particle spectral functions of symmetric nuclear matter at finite temperature for several interactions. 
We work with an in-medium $T$-matrix approach using the SCGF framework, with bare and renormalized chiral N3LO 2NF, as well as traditional phase-shift-equivalent potentials Av18 and CD-Bonn. Within the chiral approach, we also include N2LO 3NFs to identify any potential effects due to many-body interactions. By demanding that the dispersion relations of the self-energy are satisfied at every step of our self-consistent calculation, we enforce the analytical properties of the Green's functions and, formally, we expect the sum rules to be satisfied. In practice, the sum rules are a good test of the numerical consistency of the calculations. We find that the results of the zeroth, first and second moments are very well satisfied and in accordance with previous studies. We analyze, for the first time in nuclear matter, the second moment, $m^{(2)}_k$, which in turn is directly related to the variance of the spectral function. 

The first sum rule for chiral interactions has a qualitatively similar structure to the results obtained with hard forces, although there are quantitative differences. For soft interactions, the attraction at low momenta generates  binding, in contrast to other hard-core interactions like Av18. This  can be explained in terms of the generalized Hartree--Fock structure of this moment. The strength of the particle component, $m_k^>$, below the Fermi surface is directly related to the high energy tails of the spectral function. For SRG-renormalized forces, this component is negligible. The unrenormalized N3LO (or N3LO+3NF) results yield values close to $m_0^> \approx 20$ MeV, whereas harder interactions yield results in the $40-80$ MeV region. The attractive $m_k^<$ components for $k<k_F$ are also somewhat sensitive to correlations, although patterns there are more difficult to identify. 

The second moment is qualitatively different for chiral interactions and for hard forces. Whereas chiral forces, independently of the renormalization scheme, show a clear minimum as a function of $k$, harder forces provide monotonically increasing $m^{(2)}_k$. The difference in this behaviour can be partially ascribed to the variance of the spectral function, $\sigma_k^2$. The latter quantifies unambiguously the fragmentation of the quasi-particle peak, and is hence a good proxy for the softness or hardness of a nuclear interaction. We find that the variance is substantially reduced when a 2NF is renormalized using the SRG, and that the reduction is proportional to the renormalization scale. In other words, more renormalized forces have smaller variances. The bare EM N3LO force, independently of the effect of 3NF, provides a variance which, below $k<1$ GeV is $\sigma_k^2 \approx 10^4$ MeV$^2$ and which steadily decreases with momentum. Traditional hard interactions, in contrast, yield substantially larger variances, $\sigma_k^2 \approx 7 \times 10^4 - 1.5 \times 10^5$ MeV$^2$, in accordance to the increased fragmentation of the quasi-particle peak. Moreover,  $\sigma_k^2$ is an increasing function of momentum for traditional phase-shift-equivalent interactions. Both moments are sensitive to the SRG renormalization scale, as expected from their non-observable nature. We find that the inclusion of 3NFs does not reduce the scale dependence of these moments. The saturation of the sum rules as a function of energy also depends on the nuclear Hamiltonian. 

To conclude, we have found that the energy weighted sum rules of the spectral functions provide a quantitative measure of the hardness of nuclear interactions. These are directly linked to the imprints of forces in the high-energy tails of the spectral functions. In particular, the variance of the single-particle strength is particularly sensitive to the renormalization scheme and can be used as a proxy for ``perturbativeness" or ``softness" of a nuclear force. We expect that these results will be relevant for \emph{ab initio} finite nucleus calculations, where the first moment of the spectral function is already in use in the context of effective single-particle energies \cite{Duguet2012,Duguet2015}. The second moment should provide insightful quantitative information on the fragmentation of states. Its running with the SRG scale in finite nucleus should provide useful information on renormalization effects in \emph{ab initio} nuclear physics predictions.\\

\begin{acknowledgments}
This material is based upon work supported by
STFC through Grants No. ST/I005528/1, No. ST/J000051/1, No. ST/L005743/1 and No. ST/L005816/1;
 by the Deutsche Forschungsgemeinschaft through Grant No. SFB 1245; 
Grant No. FIS2014-54672-P from MICINN (Spain), and Grant No. 2014SGR-401 from Generalitat de Catalunya (Spain). 
A.C. acknowledges support by the Alexander von Humboldt Foundation through a Humboldt Research Fellowship for Postdoctoral Researchers.
Partial support comes from ``NewCompStar" COST Action MP1304. 
\end{acknowledgments}

\bibliographystyle{apsrev4-1}
\bibliography{biblio}

\end{document}